%% file: master.tex
\documentclass[letterpaper,twocolumn,10pt]{article}
\usepackage{usenix2019}

\usepackage[pdftex]{graphicx}
\graphicspath{{./Figures/}{./Figures/plots/}}

\usepackage{hyperref}
\usepackage{url}

\usepackage{amsmath}
\usepackage[utf8]{inputenc}
\usepackage{xspace}
\usepackage{hyphenat} 
\usepackage{comment}
\usepackage{booktabs}
\usepackage{subcaption}
\usepackage{multirow}
\usepackage{color, colortbl}
\usepackage{wasysym}
\usepackage{mdwlist}

\usepackage{balance}
\usepackage{soul}
\usepackage[anythingbreaks]{breakurl}
\usepackage[normalem]{ulem}
\usepackage[table,xcdraw]{xcolor}
\usepackage{rotating}
\usepackage{pgf}
\usepackage{afterpage}
\usepackage{pifont}

\newcommand{\ovs}{OvS\xspace}

\newcommand{\xmark}{\ding{55}}

\newcommand{\vs}{vswitch\xspace}
\newcommand{\vswvm}{vswitch VM\xspace}

\newcommand{\nic}{NIC\xspace}
\newcommand{\mac}{MAC\xspace}
\newcommand{\vlan}{Vlan\xspace}
\newcommand{\sriov}{SR-IOV\xspace}
\newcommand{\pf}{PF\xspace}
\newcommand{\vf}{VF\xspace}
\newcommand{\Lg}{LG\xspace}
\newcommand{\dut}{DUT\xspace}
\newcommand{\vm}{VM\xspace}
\newcommand{\os}{OS\xspace}
\newcommand{\host}{Host\xspace}
\newcommand{\baseline}{Baseline\xspace}
\newcommand{\layer}{Layer\xspace}
\newcommand{\levelone}{Level-1\xspace}
\newcommand{\leveltwo}{Level-2\xspace}
\newcommand{\levelthree}{Level-3\xspace}

\newcommand{\layertwo}{L2\xspace}
\newcommand{\cpu}{CPU\xspace}
\newcommand{\dpdk}{DPDK\xspace}
\newcommand{\ip}{IP\xspace}
\newcommand{\tcp}{TCP\xspace}
\newcommand{\udp}{UDP\xspace}
\newcommand{\apache}{Apache\xspace}
\newcommand{\memcached}{Memcached\xspace}
\newcommand{\io}{I\slash O\xspace}
\newcommand{\vxlan}{VXLAN\xspace}

\definecolor{Gray}{gray}{0.9}
\newcolumntype{w}{>{\columncolor{white}}p{1.0cm}}

\usepackage{tikz}

\newcommand{\system}{\emph{MTS}\xspace}

\def\checkmark{\tikz\fill[scale=0.25](0,.35) -- (.25,0) -- (1,.7) -- (.25,.15) -- cycle;}

\newcommand*\circled[1]{\tikz[baseline=(char.base)]{
            \node[shape=circle,draw,inner sep=1pt] (char) {#1};}}

\begin{document}
\title{Virtual Network Isolation: Are We There Yet?}
\title{Towards Network Isolation: A Secure Virtual Switch System}
\title{A Secure Virtual Switch System for Network Isolation and
Performance}
\title{\emph{MultiTenantSwitch}: Bringing Multi-Tenancy to Virtual Networking}
\title{\emph{MTS}: Bringing Multi-Tenancy to Virtual Networking}

\author{
{\rm Kashyap Thimmaraju}\\
Security in Telecommunications\\
Technische Universit{\"a}t Berlin
\and
{\rm Saad Hermak}\\
Security in Telecommunications\\
Technische Universit{\"a}t Berlin
\and
{\rm G\'abor R\'etv\'ari}\\
Information Systems Research Group\\
MTA--BME
\and
{\rm Stefan Schmid}\\
Faculty of Computer Science\\
University of Vienna
} 
\author{
{\rm Kashyap Thimmaraju$^{1}$ \quad Saad Hermak$^1$ \quad G\'abor
R\'etv\'ari$^2$\quad Stefan Schmid$^{3}$}\\
\normalsize{{$^1$ Technische Universit{\"a}t Berlin \quad
\quad $^2$ BME HSNLab
\quad~$^3$ Faculty of Computer Science, University of Vienna}}
}

%
%
%
%

\maketitle

\subsection*{Abstract}
  Multi-tenant cloud computing provides great benefits in terms of resource
sharing, elastic pricing, and scalability, however, it also changes the security
landscape and introduces the need for strong isolation between the tenants,
\emph{also inside the network}.  This paper is motivated by the observation that
while multi-tenancy is widely used in cloud computing, the virtual switch
designs currently used for network virtualization lack sufficient support for
tenant isolation.  Hence, we present, implement, and evaluate a virtual switch
architecture, $\system$, which
brings secure design best-practice to the context of multi-tenant virtual
networking: compartmentalization of virtual switches, least-privilege execution,
complete mediation of all network communication, and reducing the trusted
computing base shared between tenants. We build $\system$ from commodity
components, providing an incrementally deployable and inexpensive upgrade path
to cloud operators. Our extensive experiments, extending to both
micro-benchmarks and cloud applications, show that,
depending on the way it is deployed, $\system$ may produce 1.5-2x the throughput
compared to state-of-the-art, with similar or better latency and modest resource
overhead (1 extra \cpu{}). $\system$ is available as open source software.

\sloppy

\section{Introduction}

\begin{figure*}[t!]
    \centering
    \includegraphics[trim=0.0cm 0.0cm 0.0cm 0.0cm,clip=true]{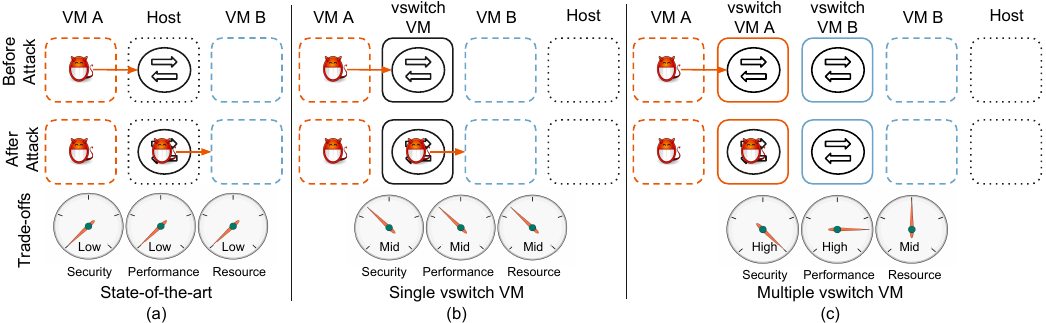}
	\caption{A high-level view of the tradeoffs between security,
performance and resources for the state-of-the-art and \system{}}
    \label{fig:motivation}
	\vspace{-1.0em}
\end{figure*}

\noindent%
\textbf{Security landscape of cloud virtual networking.} %
Datacenters have become a critical infrastructure of our digital society, and
with the fast growth of data centric applications and AI\slash ML workloads,
dependability requirements on cloud computing will further increase~\cite{gartner2017}.
At the heart of an efficiently operating datacenter
lies the idea of resource sharing and \emph{multi-tenancy}: independent
instances (e.g., applications or tenants) can utilize a given infrastructure
concurrently, including the compute, storage, networking, and management
resources deployed at the data center, in a physically integrated but logically
\emph{isolated} manner~\cite{netvirt-mtd, kube-multi-tenancy}.

At the level of the data center communication network, isolation is provided by
the network virtualization architecture.  Key to network virtualization is the
\emph{virtual switch (\vs{})}, a
network component located in the \host{} virtualization layer of the (edge) servers that
connects tenants' compute and storage resources (e.g., Virtual Machines (\vm{}s),
storage volumes, etc.), provisioned at the server, to the rest of the data center
and the public Internet~\cite{netvirt-mtd, jain2013network, ovn}.

Multi-tenancy is typically provided in this design by (i)~deploying the
\vs{}es with the server's \host{} operating system\slash hypervisor (e.g,
Open vSwitch aka \ovs{}~\cite{pfaff2015design}); (ii)~using \emph{flow-table-level
  isolation}: the \vs{}'s flow tables are divided into per-tenant
  logical datapaths that are populated with sufficient flow table entries
to link tenants' data-center-bound resources into a common interconnected
workspace \cite{netvirt-mtd,jain2013network, ovn}; and (iii) overlay networks
using a tunneling protocol, e.g., \vxlan{}~\cite{vxlanrfc}, to
  connect tenants' resources into a single workspace.  Alternatives to
this \host{}-based \vs{} model~\cite{pfaff2015design}, e.g., \nic{}-based
\vs{} solutions~\cite{baba-sriov,smartnic-broadcom} and FPGA-based
designs~\cite{accelnet}, share the main trait that the logical datapaths
have a common networking substrate (\vs{}).

Despite the wide-scale deployment~\cite{andromeda-2018,accelnet,baba-sriov}, the level of (logical and
performance)
isolation provided by \vs{}es is not yet well-understood.
For example,
Thimmaraju et al.~\cite{ovs-cloud} uncovered a serious isolation problem with a
popular virtual switch (\ovs{}). An adversary could
not only break out of the \vm{} and attack all applications on the \host{}, but could
also manifest as a worm, and compromise an entire datacenter in a few minutes.
Csikor et al.~\cite{sigcomm18policy} identified
a severe performance isolation vulnerability, also in \ovs{}, which results in a
low-resource cross-tenant denial-of-service attack.
Such attacks may exacerbate concerns surrounding the security and adoption of
public clouds (that is already a major worry across cloud users \cite{csa-survey}).

Indeed, a closer look at the cloud virtual networking best-practice, whereby
\emph{per-tenant logical datapaths are deployed on a single \host{}-based \vs{}
using flow-table-level isolation}~\cite{netvirt-mtd,jain2013network, ovn}, reveals
that the current state-of-the-art violates basically all relevant secure system
design principles~\cite{saltzer1975protection, bishop2005introduction}.  First,
the principle of \emph{least privilege} would require that any system component
should be given only the minimum set of privileges necessary to complete its
task, yet we see that \vs{} code typically executes \emph{on the \host{}} with
administrator, or what is worse, with full kernel privilege~
\cite{vmwareStart,honda2015mswitch}, even though this would not be absolutely necessary (see
Sec.~\ref{sec:system}). Second, untrusted user code directly interacts with the \vs{} and
hence with the \host{} \os{}, e.g., it may send arbitrary packets from \vm{}s, query
statistics, or even install flow table entries through side
channels~\cite{sigcomm18policy}, which violates the secure design principle of
\emph{complete mediation}. But most importantly, the shared \vs{} design goes
directly against the principle of the \emph{least common mechanism}, which would
minimize the amount of resources common to more than one tenant.

\noindent\textbf{Secure \vs{} design.} %
The main motivation for our work is the observation that \emph{current virtual
switch architectures are not well-suited for multi-tenancy}.  This observation
leads us to revisit the fundamental design of secure \vs{}es.
Hence, in this paper, we present, implement, and evaluate a multi-tenant (virtual)
switch architecture, $\system$, which extends the benefits of multi-tenancy to the
\vs{} in a secure manner, without imposing prohibitive resource requirements or
jeopardizing performance.

Fig.~\ref{fig:motivation} illustrates the key idea underlying the $\system$
design, by showing the security-performance-resource tradeoffs for different
architectures.
The current \vs{} architecture is shown in Fig.~\ref{fig:motivation}(a), whereby per-tenant
logical datapaths share a common (physical or software) switch component
deployed at the \host{} hypervisor layer (in the rest of this paper, we shall
sometimes refer to this design point as the ``\baseline{}'').  As we argued
above, this design is fundamentally insecure~\cite{ovs-cloud, sigcomm18policy}
as it violates basic secure design principles, like least privilege, complete
mediation, or the least common mechanism.
In $\system$, we address the least privilege principle by the
\emph{compartmentalization} of the \vs{}es (Fig.~\ref{fig:motivation}(b)):
by moving the \vs{}es into a dedicated \vs{} \vm{}, we can
prevent an attacker from compromising the \host{} via the \vs{}~\cite{181358}.
Then, we establish a secure communication channel between the tenant \vm{}s and the
\vs{} \vm{} via a trusted hardware technique, Single Root Input\slash Output
Virtualization, or \sriov{}, a common feature implemented in most modern \nic{}s and
motherboards \cite{amazon-sriov,baba-sriov}. Thus, all tenant-to-tenant
and tenant-\host{} networking is \emph{completely mediated} via the \sriov{} \nic{}. Adopting
Google's \emph{extra security layer} design
principle~\cite{google-securecontainer} which requires that between any
untrusted and trusted component, there have
to be at least two distinct security boundaries~\cite{lwn, kube-multi-tenancy},
we introduce a second level of isolation by moving the \vs{},
deployed into the \vs{} \vm{}, to the user space.  Hence, at least two
independent security mechanisms need to fail (user-kernel separation and
\vm{}-isolation) for the untrusted tenant code to gain access to the \host{}.

Interestingly, we are able to show the resultant secure \vs{} design, which we
call the \emph{single \vs{} \vm{}} design, does not come at the cost of
performance; just the contrary, our evaluations show that we can
considerably improve throughput and latency, for a relatively small price in
resources.  Finally, we introduce a ``hardened'' $\system$ design that we call
the \emph{multiple \vs{} \vm{}s} design (Fig.~\ref{fig:motivation}(c)),
whereby, in line with the principle of
the least common mechanism, we further separate the \vs{} by
creating multiple separate \vs{} \vm{}s, one for each tenant (or based on
security zones/classes). This way, we
can maintain full network isolation for multiple tenants.


\noindent%
\textbf{Contributions.} %
Our main contributions in this paper are:
\begin{itemize}
\item We \emph{identify requirements and design principles} that can
  prevent the virtual switch from being a liability to virtualization in
  the cloud, and we carefully apply these principles to revisit
  multi-tenancy in virtual networking.

\item We \emph{present $\system$, a secure \vs{} design} whereby the
  \vs{} is moved into a separate
  \vm{}
  that prevents malicious tenants from compromising the \host{} via the
  virtual switch, and we also show a ``hardened'' $\system$ design that
  also prevents compromising \emph{other} tenants' virtual networks through
  the \vs{}.  All our designs are incrementally deployable, providing an
  inexpensive deployment experience for cloud operators.

\item We \emph{report on extensive experiments} with our \system prototype
  and we find a noteworthy improvement (1.5-2x) in throughput compared to the
  \baseline{}, with similar or better latency for an extra \cpu{}.
  We build our prototype from
  off-the-shelf commodity components and existing software;
  \system and the data from this paper are available at:
\begin{center}
\vspace{-0.5em}
{\tt https://www.github.com/securedataplane}
\vspace{-0.5em}
\end{center}
\end{itemize}

\noindent\textbf{Organization.} %
We dive deeper into designing a secure \vs{} in
Section~\ref{sec:secure-virt-switch}. In Section~\ref{sec:system} we elaborate
on \system{} and report on two evaluations in Sections~\ref{sec:eval}
and~\ref{sec:workload}. We enter a discussion of \system{} in
Section~\ref{sec:discussion}, review related work in
Section~\ref{sec:relatedwork} and finally draw conclusions in
Section~\ref{sec:conclusion}.

\section{Securing Virtual Switches}
\label{sec:secure-virt-switch}

As demonstrated in previous work~\cite{sigcomm18policy, ovs-cloud}, the
current state-of-the-art in virtual switch design can be exploited to not only
break network isolation, but also to break out of a virtual machine.  This
motivates us to identify requirements and design principles that
make virtual switches a dependable component of the data center~\cite{secson}.

\subsection{State-of-the-Art}
\label{sec:analysis-state-art}

Virtual networks in cloud systems using virtual switches typically follow a
\emph{monolithic} architecture, where a single controller programs a \emph{single
\vs{}} running in the \host{} \os{} with per-tenant logical datapaths in the \vs{}.
Isolation between tenants is at the level of
flow-tables~\cite{netvirt-mtd,jain2013network, ovn}: the controller populates
the flow tables in each per-tenant logical datapath with sufficient flow rules
to connect the tenant's \host{}-based \vm{}s to the rest of the data center and the
public Internet. Those sets of flow rules are complex: with a small error
in one rule potentially having security consequences, e.g., making
intra-tenant traffic visible to other tenants.

As shown in Table~\ref{tab:switches}, nearly all \vs{}es are \emph{monolithic}
in nature. A single \vs{} is installed with flow rules for all the tenants
hosted on the respective server. This increases the trusted computing base (TCB)
of the single \vs{}, as it is responsible for \layer{} 2-7 of the virtual networking
stack.  Next, nearly 80\% of the surveyed \vs{}es are co-located with the \host{}
virtualization layer. This increases the TCB of the server since a \vs{} is a
complex piece of software, consisting of tens of thousands of lines of code.
The complexity of network virtualization is further increased by the fact that
packet processing for roughly 70\% of the virtual switches is spread across user
space and the kernel (see last two columns in Table~\ref{tab:switches}).  These
concerns are partially addressed by the current industry trend towards
offloading \vs{}es to \emph{smart \nic{}s}
~\cite{baba-sriov,smartnic-cavium,smartnic-broadcom,mellanox-bluefield}.
Indeed consolidating the \vs{}
into the \nic{} can improve the security as it reduces the TCB of the \host{}.  These
burgeoning architectures, however, share the main trait that the per-tenant
logical datapaths are monolithic, often with full privilege and direct access to
the \host{} \os{}, which when compromised can break network isolation and be used as a
stepping-stone to the \host{}.

\input{tables/switches.tex}

\subsection{Threat Model}
\label{sec:threat}

We assume the attacker’s goal is to either escape network virtualization by
compromising the virtual switch, or to tamper with other tenant’s network
traffic by controlling the virtual switch~\cite{ovs-cloud}. Hence, she can affordably rent a \vm{}
in an Infrastructure-as-a-Service (IaaS) cloud, or has somehow managed to
compromise a \vm{}, e.g., by exploiting a web-server
vulnerability~\cite{costin2015all}.
From the \vm{} she can send arbitrary packets, make arbitrary
computations, and store arbitrary data.  However, she does not have direct
control to configure the \host{} \os{} and hardware: all configuration access happens
through a dedicated cloud management system. 

The defender is a public cloud provider who
wants to prevent the attacker from compromising virtual network isolation; in
particular, \emph{the cloud provider wants to maintain tenant-isolation even
  when the \vs{}
  is compromised}.
We assume that the cloud provider already supports \sriov{} at
\nic{}s~\cite{amazon-sriov,baba-sriov,accelnet} and the underlying
virtualization and network infrastructure is trusted, including the
hypervisor layer, \nic{}s, firmware, drivers, core switches, and so on.

\subsection{Design Principles and Security Levels}
\label{sec:secure-design-princ}

Our \system design is based on the application of the secure system design
principles, established by Saltzer et al.~\cite{saltzer1975protection} (see also
Bishop~\cite{bishop2005introduction} and Colp et al.~\cite{colp2011breaking}),
to the problem space of virtual switches.

\noindent%
\textbf{Least privilege \vs{}.} %
The \vs{} should have the minimal privileges sufficient to complete its task, which is to
process packets to and from the tenant \vm{}s. Doing so limits the damage that can
result from a system compromise or mis-configuration.
Current best-practice is, however, to run the \vs{} co-located with the
\host{} \os{} and with elevated privileges; prior work has shown the types
and severity of attacks that can happen when this principle fails
\cite{ovs-cloud}.
A well-known means to the principle of least privilege is
\emph{compartmentalization}: execute the \vs{} in an isolated environment with
limited privileges and minimal access to the rest of the system.
In the next section, we will show how \system implements compartmentalization by
committing the \vs{}es into one or more dedicated \vs{} \vm{}s.

\noindent%
\textbf{Complete mediation of tenant-to-tenant and tenant-to-host networking.} %
This principle requires that the network communication between the untrusted
tenants and the trusted \host{} is completely mediated by a trusted intermediary to prevent
undesired communication. This principle, when systematically applied, may go a
long way towards reducing the \vs{} attack surface.
By \emph{channeling all network communication} between untrusted and trusted
components \emph{via a trusted intermediary} (a so called reference monitor),
the communication can be validated, monitored and logged based on security
policies. In the next section, we show how complete mediation is realized in
\system using a secure \sriov{} channel between the tenant \vm{}s, \vs{}es and
\host{}.

\noindent%
\textbf{Extra security boundary between the tenant and the host.} %
This security principle, widely deployed at
Google~\cite{google-securecontainer}, requires
that between any untrusted and trusted component there has to be at least two
distinct security boundaries, so at least two independent security mechanisms
need to fail for the untrusted component to gain access to the
sensitive component~\cite{lwn}. We establish this extra layer of security in
\system by \emph{moving the \vs{} to user space}.  This
also contributes to implementing the ``least privilege'' principle: the
user-space \vs{} can drop administrator privileges after
initialization.

\noindent%
\textbf{Least common mechanisms.} %
This principle addresses the amount of infrastructure shared between
tenants; applied to the context of \vs{}es this principle requires that
the network resources (code paths, configuration, caches) common to more
than one tenant should be minimized. Indeed, every shared resource may
become a covert channel~\cite{bishop2005introduction}. Correspondingly,
\emph{decomposing the \vs{}es themselves into multiple compartments}
could lead to hardened vswitch designs.

\noindent%
\textbf{Security levels.} %
From these principles, we can obtain different levels of security:
\begin{itemize}
\item \textbf{\baseline{}:} The per-tenant logical datapaths are consolidated into a single
physical or software \vs{} that is co-located with the \host{} \os{}.
\item \textbf{\levelone{}:} Placing the \vs{} in a dedicated compartment provides a
first level of security by protecting from malicious tenants to compromise the
\host{} \os{} via the \vs{} (``single \vs{} \vm{}'' in Fig.~\ref{fig:motivation}b).
\item \textbf{\leveltwo{}:}
Splitting the \vs{}es into multiple compartments (based on security zones or on
a per-tenant basis) adds another level of security, by isolating tenants' \vs{}es
from each other (``multiple \vs{} \vm{}s'' in Fig.~\ref{fig:motivation}c).
\item \textbf{\levelthree{}:} Moving the \vs{}es into user space,
combined with \baseline{} or \levelone{} or -2, reduces the impact of a compromise and further
reduces the attack surface. 
\end{itemize}

\section{The \system Architecture}
\label{sec:system}

We designed \system with secure design principles from
Section~\ref{sec:secure-design-princ}.
We first provide an overview and then present our architecture
in detail.

\subsection{Overview}

\noindent\textbf{Compartmentalization.} %
There are many ways in which isolated \vs{} compartments can be implemented:
full-blown \vm{}s, \os{}-level sandboxes (jails, zones, containers,
plain-old user-space processes~\cite{kube-multi-tenancy}, and exotic
combinations of these \cite{kata-containers, google-gvisor}), hardware-supported
enclaves (Intel's SGX)~\cite{paladi2019sdn, 199364}, or even safe
programming language compilers (Rust), runtimes (eBPF), and instruction sets
(Intel MPX). For flexibility, simplicity, and ease of deployment,
\system \emph{relies on conventional \vm{}s as the main unit of
  compartmentalization}.

\begin{figure}[t!]
    \centering
    \includegraphics[trim=0.0cm 0.0cm 0.0cm 0.0cm,clip=true,width=.99\columnwidth]{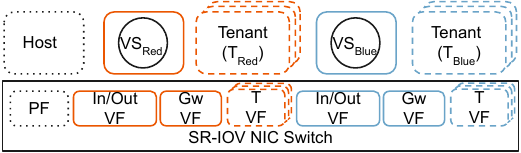}
	\caption{High-level overview of \system in security \leveltwo{}. The Red and
Blue \vs{} compartments (\vm{}s) are allocated dedicated virtual functions (\vf{}s)
to communicate with external networks using  the \emph{In\slash Out \vf{}}, their
respective tenants using the \emph{Gw \vf{}} and \emph{T \vf{}}.
Communication between the \vs{}es, tenants and the \host{} physical
function (\pf{}) are mediated via the \sriov{} \nic{} switch. }
    \label{fig:system}
    \vspace{-1.0em}
\end{figure}

\vm{}s provide a rather strong level of isolation and are widely supported in
hardware, software, and management systems. This in no way means that
\vm{}-based \vs{}es are mandatory for \system, just that this approach
offers the highest flexibility for prototyping. For simplicity,
Fig.~\ref{fig:system} depicts two \vs{} compartments (Red and Blue
solid boxes) running independent \vs{}es in their isolated \vm{}s. The
multiple compartments further reduce the common mechanisms between the
\vs{} and the connected tenants, achieving security \leveltwo{}. Security
\levelone{}, although not depicted, would involve only a single \vs{} \vm{}.

\noindent%
\textbf{Complete mediation.} %
To mediate all interactions between untrusted tenant code and the \host{}
\os{} through the \vs{}, we need a secure and high-performance communication
medium between the corresponding compartments\slash \vm{}s.  In
\system \emph{we use Single Root IO Virtualization (\sriov{}) to
inter-connect the \vs{} compartments} (see Figure~\ref{fig:system}).

\sriov{} is a PCI-SIG standard to make a single PCIe device, e.g., a \nic{},
appear as multiple PCIe devices that can then be attached to multiple \vm{}s.
An \sriov{} device has one or more physical functions (\pf{}s) and one or more virtual
functions (\vf{}s), where the \pf{}s are typically attached to the \host{} and the \vf{}s to
the \vm{}s.
Only the \host{} \os{} driver has privileges to configure the \pf{}s and \vf{}s. 
The \nic{} driver in the \vm{}s in turn have restricted access to \vf{} configuration.
Only via the \host{}, \vf{}s and \pf{}s can be configured with unique \mac{} addresses and \vlan{} tags.
Network communication between the \pf{}s and \vf{}s occurs via an \layertwo{}
switch implemented in the \nic{} based on the IEEE Virtual Ethernet Bridging
standard~\cite{veb}. This enables Ethernet
communication not only from and to the respective \vm{}s (\vs{} and tenants)
based on the destination \vf{}'s \mac{} address but also to the external networks.

Sharing the \nic{} \sriov{} \vf{} driver and the \layer{} 2 network virtualization
mechanism implemented by the \sriov{} \nic{}
is considerably simpler than including the \nic{} driver and the entire network
virtualization stack (\layer{} 2-7) in the TCB. Tenants already share
SR-IOV NIC drivers in public clouds~\cite{amazon-sriov,baba-sriov,azure-sriov}.
Virtual networks can be built as we will see next, as per-tenant user-space
applications implementing \layer{} 3-7 of the virtual networking stack.

Thanks to the use of \sriov{} in \system, packets to and from tenant \vm{}s completely
bypass the \host{} \os{}; instead, all potentially malicious traffic is channeled
through the trusted hardware medium (\sriov{} \nic{}) to the \vs{} \vm{}(s).
Furthermore, using \sriov{} reduces \cpu{} overhead and improves performance
(see Section~\ref{sec:eval}).
Finally, \sriov{} provides an attractive upgrade path towards fully
offloaded, smart-\nic{} based virtual networking: chip~\cite{kutch2011pci}
and \os{} vendors~\cite{linux-sriov, bsd-sriov} have
been supporting \sriov{} for many years now at a reasonable price, major cloud
providers already have \sriov{} \nic{}s deployed in their data
centers~\cite{amazon-sriov,baba-sriov,azure-sriov}, and, perhaps
most importantly, this design choice liberates us from having to re-implement
low-level and complex network components~\cite{181358}: we can simply use any desired \vs{},
deploy it into a \vs{} \vm{}, configure and attach \vf{}s to route tenants'
traffic through the \vs{}, and start processing packets right away.

\noindent%
\textbf{User-space packet processing.} %
As discussed previously, we may choose to deploy the \vs{}es into the
\vs{} \vm{} user-space to establish an extra security boundary between
the tenant and the \host{} \os{} (\levelthree{} design).  Thanks to the advances in kernel
bypass techniques, several high-performance and feature-rich user-space packet
processing frameworks are available today, such as Netmap~\cite{rizzo2012netmap},
FD.IO~\cite{fdio}, or Intel's \dpdk{}~\cite{dpdk}.
Our current design of \system \emph{leverages \ovs{} with the \dpdk{}
datapath for implementing the \vs{}es}~\cite{dpdkOvs}.
\dpdk{} is widely supported, it has already been integrated with
popular virtual switch products, and extensive operational experience is
available regarding the expected performance and resource footprint
\cite{8468219}.  Note, however, that using \dpdk{} and \ovs is not mandatory in
\system; in fact, thanks to the flexibility provided by our \vm{}-compartments and
\sriov{}, we can deploy essentially any user-space \vs{} to support \system.

\subsection{Detailed Architecture}
\label{sec:deta-arch}

\begin{figure*}[t!]
    \centering
    \includegraphics[trim=0.0cm 0.0cm 0.0cm 0.0cm,clip=true,width=.99\textwidth]{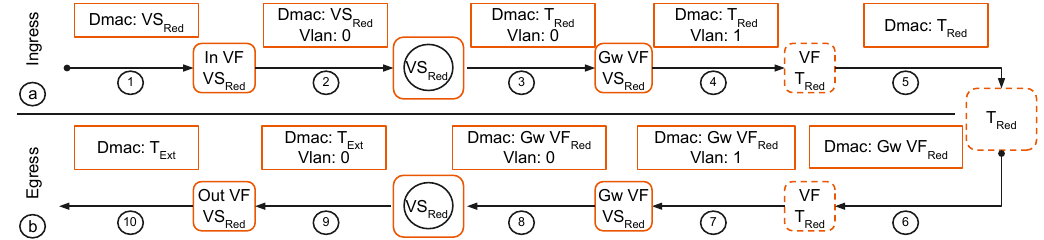}
	\caption{A step-by-step illustration of how packets enter and leave the
Red tenant from Figure~\ref{fig:system} in \system.
\protect\circled{a} shows how ingress packets
reach Tenant$_{Red}$. \protect\circled{b} shows how Tenant$_{Red}$ packets reach an
external system Tenant$_{Ext}$.}
    \label{fig:systemDetail}
    \vspace{-1.0em}
\end{figure*}

For the below discussion, we consider the operation of \system for one \vs{}
compartment and its corresponding tenant \vm{}s from the \leveltwo{} design shown in
Fig.~\ref{fig:system}.
The case when only a single compartment (\levelone{}) is used is similar in vein:
the flow table entries installed into the \vs{} and the \vf{}s attached to the
\vs{} compartment need to be modified somewhat; for lack of space we do not
detail the \levelone{} design any further.

\noindent\textbf{Connectivity.} %
Each \vs{} \vm{} is allocated at least one \vf{} (In\slash Out \vf{}) for external (inter-server) connectivity
and another as a gateway (Gw \vf{}) for \vs{}-\vm{}-to-tenant-\vm{}
connectivity as shown in Fig.~\ref{fig:system}.  Isolation between the
external and the tenant network (tenant \vf{} shown as T \vf{} in the Figure) is enforced at the \nic{}-level by
configuring the Gw \vf{} and the tenant \vf{}s with a \vlan{} tag specific to the tenant.
Different \vlan{} tags are used to further isolate the multiple
\vs{} compartments and their resp.~tenants on that server.

The packets between \vf{}s\slash \pf{}s in the \nic{} are forwarded based on the
destination \mac{} address and securely isolated using \vlan{} tags (the same
security model as provided by enterprise Ethernet switches). For all
packets to and from the tenant \vm{}s to pass through the \vs{}-\vm{}, the
destination \mac{} address of each packet entering and leaving the \nic{} needs to be
accurately set, otherwise  packets will not reach the correct destination.  This
can be addressed by introducing minor configuration changes to the normal
operation of the tenant and the \vs{}es, detailed below.

\noindent\textbf{Ingress chain.} %
Fig.~\ref{fig:systemDetail}\circled{a} illustrates the process by which
packets from an external network reach the tenant \vm{}s.  In step \circled{1} a packet enters the server through
the \nic{} fabric port having the Red \vs{}'s In\slash Out \vf{} \mac{} address as the
destination \mac{} address (Dmac). The \nic{} switch will deliver the packet to the
\vs{} \vm{} untagged (\vlan{} 0) in~\circled{2}. The Red \vs{} then
uses the destination IP address in the packet to identify the correct tenant \vm{}
to send the packet to, changes the destination \mac{} address to that of the Red
tenant's \vf{} (\vf{} $T_{Red}$), and emits the packet to the Gw \vf{} in the \nic{} in
\circled{3}. This ensures accurate packet delivery to and from tenant \vm{}s and
the complete isolation of the tenant-\vs{} traffic from other traffic
instances. In \circled{4} and \circled{5}, the \nic{} tags the packet with the
Red tenant's specific \vlan{} tag (\vlan{} 1 in the figure), uses the built-in switch
functionality to make a lookup in the \mac{} learning table for the \vlan{}, pops
the \vlan{} tag and finally forwards the packet to the Red tenant's \vm{}.
The \nic{} forwarding process is completely transparent to the
\vs{} and tenant \vm{}s, the only downside is the extra round-trip to the \nic{}.
Later we show that this round-trip introduces negligible latency overhead.

\noindent\textbf{Egress chain.} %
The reverse direction shown in Fig.~\ref{fig:systemDetail}\circled{b},
sending a packet from the tenant \vm{} through the \vs{} to the external
network goes in similar vein. In \circled{6} the Red tenant \vm{} $T_{Red}$
sends a packet through its \vf{} ($T_{Red}$) with the destination \mac{} address
set to the \mac{} address of the Red tenant's Gw \vf{}; in the next subsection
we describe two ways to achieve this.  In \circled{7} the \nic{} switch
tags the packet (\vlan{} 1), looks-up the destination \mac{} address which
results in sending the packet to the Gw \vf{}. At the gateway \vf{} \circled{8},
the \nic{} switch pops the \vlan{} tag and delivers the packet to the Red \vs{}
\vm{}. The \vs{} receives the packet, looks up the destination IP address,
rewrites the \mac{} address to the actual (external) gateway's \mac{} address,
and then sends the packet out to the In\slash Out \vf{} in
\circled{9}. Finally in \circled{10}, the \nic{} will in turn send the packet
out the physical fabric port.


Communication between the two \vm{}s of a single tenant inside the server goes
similarly, with the additional complexity that packets now take \emph{two} extra
round-trips to the \nic{}: once on the way from the sender \vm{} to \vs{}, and once
on the way from the \vs{} to the destination \vm{}. Again, our evaluations in the
next sections will show that the induced latency overhead for such a traffic
scenario is low.

\noindent%
\textbf{System support.} %
Next, we detail the modifications the cloud operator needs to apply to the
conventional \vs{} setup to support \system.
The primary requirement is to modify the centralized controllers to
appropriately configure tenant specific \vf{}s with \vlan{} tags and \mac{} addresses,
and insert correct flow rules to ensure the \vs{}-tenant connectivity.
Second, advanced multi-tenant cloud systems rely on tunneling protocols to
support \layertwo{} virtual networks. This is also supported by \system, by modifying the
flow tables to pop\slash insert the appropriate headers whenever packets need to
be decapsulated\slash encapsulated. Note that after
decapsulation the tunnel id can be used in conjunction with the destination IP
address to identify the appropriate tenant \vm{}.
Third, the ARP entry for the default gateway must be appropriately set
in each tenant \vm{} so that packets from the tenant \vm{} go to the \vs{} \vm{}. To
this end, the tenant \vm{}s can be configured with a static ARP entry pointing to
the appropriate Gw \vf{}, or using the centralized controller and \vs{} as a
proxy-ARP\slash ARP-responder~\cite{ovs-proxyarp}.
Finally, to prevent malicious tenants from launching an attack on the
system, the cloud operator needs to deploy security filters in the \nic{}.
In particular, source \mac{} address spoofing prevention must be enabled on all
tenant \vm{}s' \vf{}s. Furthermore, flow-based wildcard filters can also be applied in
the \nic{} for additional security, e.g., to drop packets not destined to the
\vs{} compartment, to prevent the \host{} from receiving packets from the
tenant \vm{}s, etc.  Our \system implementation, described in
Section~\ref{sec:eval}, takes care of removing
the manual management burden in applying the above steps.

\noindent%
\textbf{Resource allocation.} %
Additional levels of security usually come with increased resource requirement,
needed to run the security\slash isolation infrastructure.  Below, we describe
two resource sharing strategies and how the \vf{}s are allocated to the \vs{}
compartments. However, due to the sheer quantity and diversity in cloud setups,
we restrict the
discussion to plain compute and memory resources and the number of \sriov{} \vf{}s
for the different \system security levels.

We consider two modes for compute and memory resources.  A \emph{shared}
mode where tenants' \vs{}es share a single physical \cpu{} core, while in
the \emph{isolated} mode each tenant's \vs{} is pinned to a different
core.
However, we assume that each \vs{} compartment gets an
equal share of main memory (ram) and this is inexpensive compared to
physical \cpu{} cores. Dedicating compute and memory resources for \vs{}ing is
not uncommon among cloud operators~\cite{accelnet,andromeda-2018}.
Note that the \emph{shared} and \emph{isolated} resource allocations are merely
two ends of the resource allocation spectrum, different sets of \vs{} \vm{}s
could be allocated resources differently, e.g., based on application or customer
requirements.
In the next section we will
see that the resource requirement for multiple \vs{} \vm{} compartments,
i.e., \leveltwo{} alone, is not resource prohibitive in the
\emph{shared} mode, however, \leveltwo{} and \levelthree{}
can be.

Regarding the number of \sriov{} \vf{}s needed,  the current standard allows
each \sriov{} device to have up to 64 \vf{}s per \pf{}.
For \levelone{}, the total number of \vf{}s is given by the sum of i) the
number of \vf{}s allocated for external connectivity (In\slash Out \vf{}); ii) the
total number of tenant-specific gateway \vf{}s; and  iii) tenant-specific
\vm{} \vf{}s hosted on the server. In a basic \levelone{} setup hosting 1 tenant, with 1
In\slash Out \vf{} and 1 gateway \vf{} and 1 \vf{} for the tenant \vm{}, the total \vf{}s is 3.
Similarly for 4 tenants, the total \vf{}s is 9.
For \leveltwo{}, the total number of \vf{}s is given by the sum of i) the tenant-specific
\vf{}s allocated for external connectivity; ii) the tenant-specific gateway
\vf{}s; and iii) tenant-specific \vm{} \vf{}s hosted on the server. For a basic \leveltwo{}
setup hosting 2 tenants, with 1 In\slash Out \vf{}, 1 gateway \vf{} per tenant \vs{}
and 1 \vf{} for each tenant \vm{}, the total \vf{}s is 6. Similarly for 4 tenants, the
total \vf{}s is 12.
%




\section{Evaluating Tradeoffs}
\label{sec:eval}



We designed a set of experiments to empirically
evaluate the security-performance-resource tradeoff
of \system. To this end, we measure  \system{}'s performance
for different security levels under different resource allocation strategies,
in canonical cloud traffic scenarios~\cite{emmerich2014performance}.
The focus is on throughput and latency performance
metrics, and physical cores and memory for resources.
In particular, the experiments serve to verify our expectation
that our design does not introduce a considerable overhead in
performance. However, we do expect the amount of resources consumed to increase; our aim is to quantify this increase in different realistic setups. 

\noindent\textbf{Prototype framework. }
We took a programmatic approach to our design and evaluation, hence, we developed a set of
primitives that can be composed to configure \system to conduct all the
experiments described in this
paper. 
Hence, as a first step we do not consider complex cloud management
systems (CMS) such as OpenStack; this way we can conduct self-contained experiments without the possible interference cause by a CMS. Our framework is written in Python and currently
supports \ovs{} and ovs-\dpdk{} as the base virtual switch, Mellanox \nic{}, and the \texttt{libvirt} virtualization framework.
Our framework and data are available on-line at the following URL:
{\tt https://www.github.com/securedataplane}



\begin{figure}[t!]
    \centering
    \includegraphics[trim=0.0cm 0.6cm 0.0cm 0.0cm,clip=true,width=.99\columnwidth]{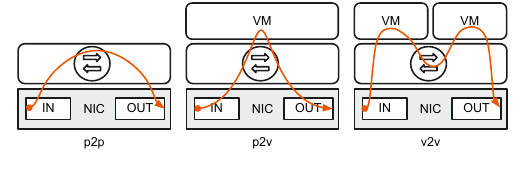}
    \caption{Traffic scenarios evaluated.}
    \label{fig:topologies}
    \vspace{-1.0em}
\end{figure}

\noindent\textbf{Methodology. }
We chose a set of standard cloud traffic scenarios (see
Fig.~\ref{fig:topologies}) and a fixed number of tenants (4). For each of
those scenarios, we allocated the necessary resources (Sec.~\ref{sec:system}) and then
configured the \vs{} either in its default configuration (\baseline{}) or one of
the three security levels (Sec.~\ref{sec:secure-design-princ}).
The system was then connected in a measurement setup to measure
the one-way forwarding performance.  
Important details on the topology, resources, security levels and the hardware and
software used are described next.

\noindent\textbf{Traffic scenarios. } 
The three scenarios evaluated are shown in Fig.~\ref{fig:topologies}.
\emph{Physical-to-physical (p2p):} Packets are forwarded by the \vs{} from the
ingress physical port to the egress. This is meant to shed light on basic \vs{}
forwarding performance.
\emph{Physical-to-virtual (p2v):} Packets are forwarded by the \vs{} from one physical
port to a tenant \vm{}, and then back from the tenant \vm{} to the other physical
port. Compared to the p2p, this will show the overhead to forward to and from the
tenant \vm{}.
\emph{Virtual-to-virtual (v2v):} Similar to the p2v, however, when the packets return from the
tenant to the \vs{}, the \vs{} sends the packet to another tenant which then
sends it back to the \vs{} and then out the egress port. This scenario emulates
service chains in network function virtualization.
Since the path length increases from p2p to p2v to v2v, we expect the
latency to increase and the throughput to decrease when going from
p2p to p2v to v2v.

\noindent\textbf{Resources. }
We allocated compute resources in the following two ways.
\emph{Shared:} All \vs{} compartments share 1 physical \cpu{} core and their
associated cache levels.
\emph{Isolated:} Each \vs{} compartment is allocated 1 physical \cpu{} core and
their associated cache levels. In case of the \baseline{}, we allocated cores
proportional to the number of \vs{} compartments, e.g., 2 cores to compare with
2 \vswvm{}s.
For main memory, each \vm{} (\vs{} and tenant) was allocated 4 GB of which 1 GB is reserved as
one 1GB Huge page. Similarly, for the \baseline{}, a proportional amount of
Huge pages was allocated.
When using \system{}, each \vswvm{} was allocated 2 In\slash Out \vf{}s
(1 per physical port), and 2 appropriately \vlan{} tagged Gw \vf{}s per tenant (1 per
physical port).
When \dpdk{} was used in \levelthree{}: one physical core needs to allocated for each
ovs-\dpdk{} compartment (including the \baseline{}), hence, only the isolated mode was
used; all In\slash Out, gateway and tenant ports connected to \ovs{} were
assigned \dpdk{} ports (in the case of the \baseline{}, the tenant port type was the
dpdkvhostuserclient~\cite{dpdkvhost}).  All the tenant \vm{}s got two physical cores and two \vf{}s, 1
per port (these are \vm{}s the tenant would use to run her application) so that the
forwarding app ($l2fwd$) could run without being a bottleneck.

\noindent\textbf{Security levels and tenants. }
For each resource allocation mode, we configured our setup either in \baseline{} or one
of the three \system security levels (Section~\ref{sec:secure-design-princ}).
In the \baseline{} and \levelone{}, there were 4 tenant \vm{}s connected to the \vs{}.
For \leveltwo{}, we configured 2 \vs{} \vm{}s and each \vs{} had 2 tenant \vm{}s, and
then we configured 4 \vs{} \vm{}s where each \vs{} \vm{} had 1 tenant \vm{}.
We repeated \levelthree{} with \baseline{}, \levelone{} and the two \leveltwo{} configurations.

\noindent \textbf{Setup. } 
To accurately measure the one-way forwarding performance (throughput and
latency), we used two servers connected to each other via 10G short range
optical links.
The device under test (\dut{}) server was an Intel(R) Xeon(R) \cpu{} E5-2683 v4 @
2.10GHz with 64 GB of RAM with the IOMMU enabled but hyper-threading and energy efficiency disabled,
and a 2x10G Mellanox ConnectX4-LN \nic{} with adaptive interrupt moderation and
irq balancing disabled.
The other server was the packet\slash load generator (\Lg{}), sink  and monitor,
with an Endace Dag 10X4-P card
(which gives us accurate and precise hardware timestamps)~\cite{endace}.
The link between the \Lg{} and \dut{}, and \dut{} and sink were
monitored via a passive optical network tap connected to the Dag card.
Each receive stream of the Dag card was allocated 4 GB to receive packets.
The \host{}, \vswvm{} and tenant \vm{}s used the Linux kernel 4.4.0-116-generic,
Mellanox OFED linux driver 4.3-1.0.1.0, \ovs{}-2.9.0 and \dpdk{} 17.11.
Libvirt 1.3.1 was used with QEMU 2.5.0. In the tenant \vm{}s, we adapted the
\dpdk{}-17.11 $l2fwd$ app to rewrite the correct destination \mac{} address when using
\system{}, and used the default $l2fwd$ drain-interval (100 microseconds)
and burst size (32) parameters.
For the \baseline{}, we used the default linux bridge in the tenant \vm{}s as
using \dpdk{} in the tenant without being backed by QEMU and \ovs{} (e.g.,
dpdkvhostuserclient) is not a recommended configuration~\cite{mellanox}.
For network performance measurements, we used Endace dag-5.6.0 software
tools (dagflood, dagbits, and dagsnap).

\subsection{Throughput}
\label{sec:tput}
Our first performance tradeoff is evaluating the forwarding throughput.
This will shed light on the packets per second (pps) processing performance of
\system{} compared to the \baseline{}. It also uncovers packet loss sooner than
measuring the bandwidth~\cite{jacobson1988congestion}. We measure the
aggregate throughput with a constant stream of 64 $B$ packets replayed at line
rate (14 Mpps) by the \Lg{} and collected at the sink. Since we fixed
the number of tenants to 4, the stream of packets comprises 4 flows, each to
a respective tenant \vm{} identified by the destination \mac{} and \ip{} address.
At the monitor we collect the packets forwarded to report the
aggregate throughput. Each experimental run lasts for 110 seconds and
measurements are made from the 10-100 second marks.

\begin{figure*}[t!]
	\centering
	\includegraphics[trim=.100cm 0.32cm .30cm 0.0cm,clip=true,height=.2\textwidth, width=1.00\textwidth]{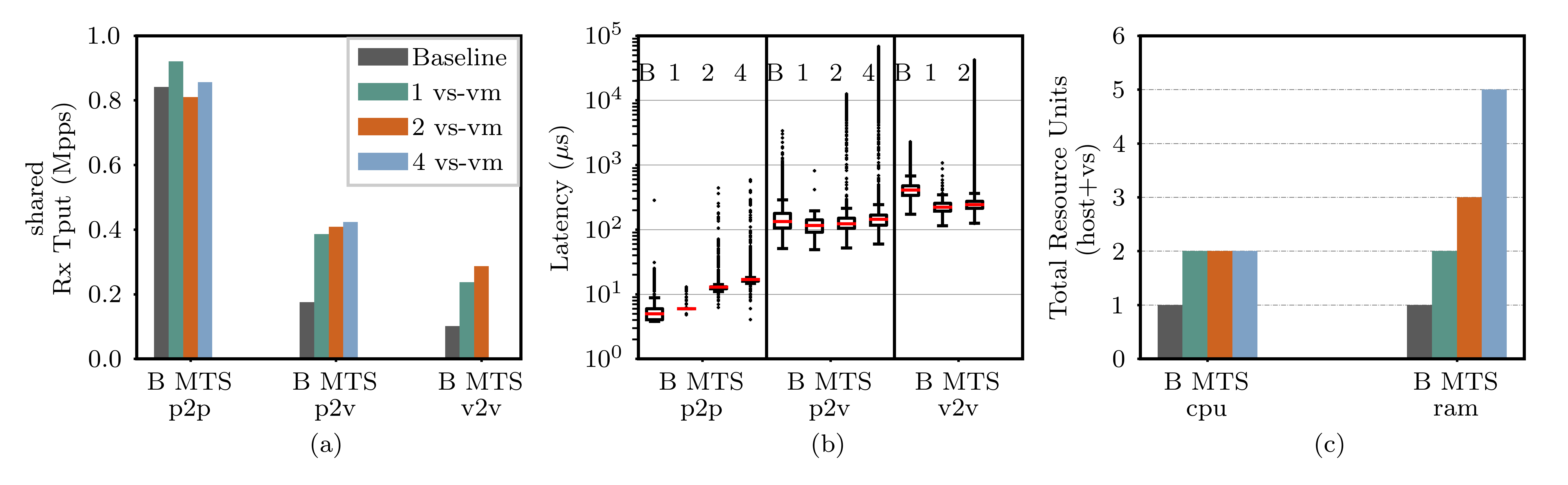}
	\includegraphics[trim=.100cm 0.32cm .30cm 0.0cm,clip=true,height=.2\textwidth, width=1.00\textwidth]{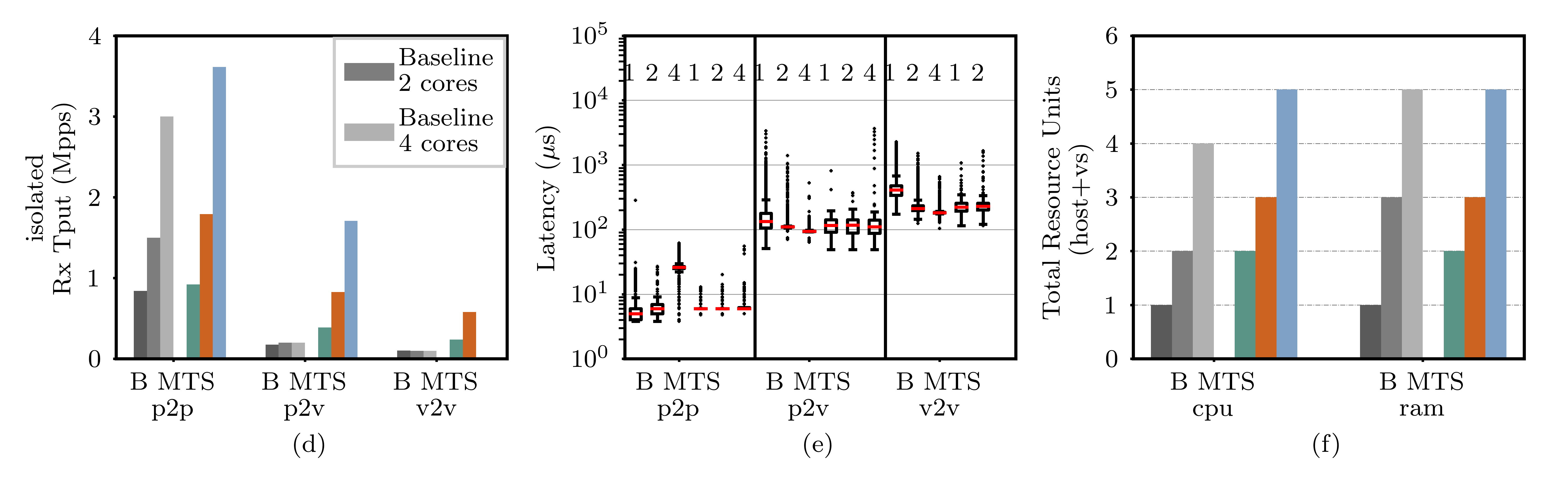}
	\includegraphics[trim=.100cm 0.32cm .30cm 0.0cm,clip=true,height=.2\textwidth, width=1.00\textwidth]{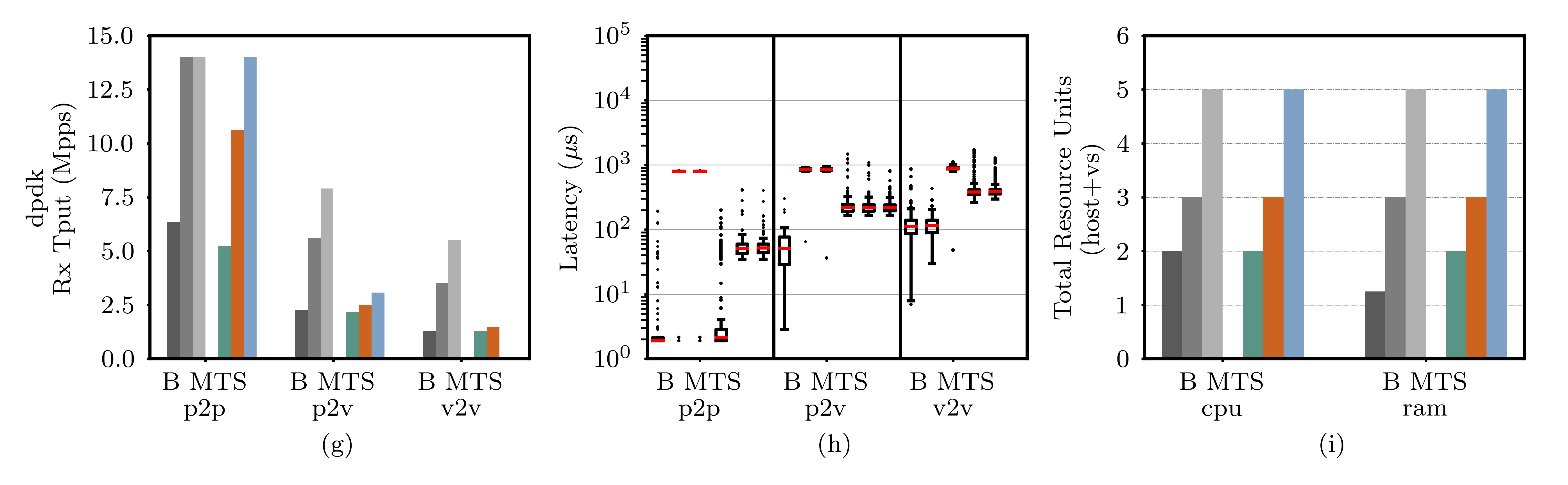}
	\caption{The security, throughput, latency and resource tradeoff comparison
of \system. The rows indicate the resource mode. The columns are ordered as
throughput, latency and resources. The security levels used are shown in the
legend. Note the bottom row is for security \levelthree{} in the isolated
resource mode combined with other security levels.}
	\label{fig:eval}
    \vspace{-1.0em}
\end{figure*}



\noindent\textbf{Results. }
The throughput measurement data for the shared mode is shown in
Fig.~\ref{fig:eval}(a). In Fig.~\ref{fig:eval}(d) we can see the data for
the isolated mode and in Fig.~\ref{fig:eval}(g) the data for \levelthree{} in the
isolated mode is shown.
From Figures~\ref{fig:eval}(a) and (d) we can see that nearly always \emph{\system{} had either the same or higher aggregate
throughput than the \baseline{}}. The improvement in throughput is most obvious in
the p2v and v2v topologies as \vs{}-to-tenant communication is via the PCIe bus
and \nic{} switch, which turns out to be faster than \baseline{}'s memory bus and software approach.
Sharing the physical core for multiple compartments (Fig.~\ref{fig:eval}(a)) in
the p2v and v2v scenarios can offer 4x isolation (\leveltwo{} with 4 compartments)
and a 2x increase in throughput (nearly .4 Mpps and .2 Mpps) compared to the
\baseline{} (nearly .2 Mpps and .1 Mpps).

Fig.~\ref{fig:eval}(d) is noteworthy as multiple cores for \vs{} \vm{}s and the
\baseline{} functions as a load-balancer when isolating the \cpu{} cores.  In the p2p scenario, the
aggregate throughput increases roughly from 1 Mpps to 2 Mpps to 4 Mpps as the
number of cores increase. We observe that \system{} is slightly more than the
\baseline{} in the p2p, however, in the p2v and v2v scenarios \system{} offers
higher aggregate throughput.
As expected, using \dpdk{} can offer an order of magnitude better throughput
(Fig.~\ref{fig:eval}(g)). In the p2p topology, we were able to nearly
reach line rate (14.4 Mpps) with four \dpdk{} compartments as the packets were load-balanced
across the multiple \vs{} \vm{}s, while the \baseline{} was able to saturate the
link with 2 cores.
With \system{}, the throughput saturates (at around 2.3 Mpps) in the p2v and v2v
topologies because several ports are polled using a single core and packets have
to bounce off the NIC twice as much compared to the Baseline where we observe
nearly twice the throughput for 2 and 4 cores.
Nevertheless, we can see a slight increase in the
throughput of \system{} as the \vs{} \vm{}s increase, because the number of ports per
\vs{} \vm{} decreases as the number of \vs{} \vm{}s increase.
Due to the limited physical cores on the \dut{}, we could not evaluate
4 \vs{} \vm{}s in the v2v topology as it required more cores and
ram than available.

\noindent\textbf{Key findings. }
The key result here is that \system{} offers increasing levels of security with
comparable, if not increasing levels of throughput in the shared and isolated
resource modes, however, the \baseline{}'s throughput with user-space packet
processing (\dpdk{}) is better than \system{}.

\subsection{Latency}
\label{sec:lat}
The second performance tradeoff we evaluated was the forwarding latency, in
particular, we studied the impact of packet size on forwarding. We selected
64$B$ (minimum IPv4 UDP packet size), 512$B$ (average packet), 1500$B$ (maximum
MTU) packets and 2048$B$ packets (small jumbo frame). As in the throughput
experiments, we used 4 flows, one to each tenant. For each experimental run, we
continuously sent packets from the \Lg{} to the sink via the \dut{} at
10 kpps for 30 seconds. Note that is the aggregate throughput sent to the \nic{}
and not to the \vswvm{}. To eliminate possible warm-up effects,
we only evaluated the packets from the 10-20 second mark.

\noindent\textbf{Results. }
For brevity the latency distribution only for 64 $B$ packets is reported here.
Fig.~\ref{fig:eval}(b) shows the data for the shared mode, while
Fig.~\ref{fig:eval}(e) is for the isolated mode. \levelthree{} latency data is shown
in Fig.~\ref{fig:eval}(h). Although the p2p scenarios
shows that \system{} increases the latency (Fig.~\ref{fig:eval}(b),
(e) and (h)), the p2v and v2v scenarios show that \system{} is slightly
faster than the \baseline{}. This is for two reasons.  First, packets between the
\vs{} and the tenant \vm{}s pass through the \sriov{} \nic{} (PCIe bus) rather than a software only
\vs{} (memory bus).  Second, when using the \baseline{} the tenant uses the Linux bridge.
The exception to this can be seen with user-space packet processing
(Fig.~\ref{fig:eval}(h)), where the \baseline{} with a single core for dpdk (2 in total) is always faster than
\system{}. As mentioned in Section.~\ref{sec:tput}, due to resource limitations
we could not evaluate the 4 \vs{} \vm{}s in v2v.

The variance in latency increases as more compartments share the same physical
core (Fig.~\ref{fig:eval}(b)). Isolating the \vs{} \vm{} cores leads to more
predictable latency as seen in Fig.~\ref{fig:eval}(e). When using \dpdk{}
(Fig.~\ref{fig:eval}(h)) we make two observations: i) \system{} takes longer to forward packets
than without using \dpdk{}; ii) the latency for \baseline{} with 2 and 4 cores
for dpdk (3 and 5 in total) is
unexpectedly high (around 1 ms). Regarding the former, we conclude that \system{} with \ovs{} and \dpdk{} requires
further tuning as we used the default \ovs{}-\dpdk{} parameters for the drain interval, batch
size and huge pages: There is an inherent
tradeoff between high throughput and average per-packet latency when using a
shared memory model where a core is constantly polling~\cite{dpdk-tput-lat}.
For the latter, we observe that the throughput of 10 kpps is too low to drain
the multiple queues on the \dpdk{} ports. At 100 kpps and 1 Mpps, we measured
an approximately 2 microsecond latency for the p2p scenario.


\noindent\textbf{Key findings. }
We observe that for the shared mode, and 4x
compartmentalization (\leveltwo{}), the latency is comparable to the \baseline{} (p2v)
with a lot of variance whereas when isolated the latency is more predictable.


\subsection{Resources}
\label{sec:resources}

In Fig.~\ref{fig:eval}(c), (f) and (i)
we see the total \cpu{} and memory consumption for \baseline{} and \system{}. Note
that across all the figures, one core and at least one Huge page is always
dedicated for the \host{} \os{}. In the case of the (single core) \baseline{}, the \vs{}
(\ovs{}) runs in the \host{} \os{} and hence shares the \host{}'s core and ram. However,
for the single \vswvm{} in the shared, isolated and \dpdk{} modes, the \host{} \os{}
consumes one core and the \vswvm{} consumes another core making the total \cpu{}
cores two. Similarly, the 2 and 4 \vswvm{}s in the shared mode, also consume the
same number of cores as the single \vswvm{} but a linear increase in ram.  In
the isolated mode, \system{} consumes only one extra physical core relative to
the \baseline{}, and in \dpdk{}, \system{} and \baseline{} consume equal number of cores.
With respect to the memory consumption, we note that \system{}'s and \baseline{}'s
memory consumption in the isolated and \dpdk{} modes are the same.

Hence, we conclude that for one extra physical core, \system{} offers multiple
compartments, making the shared resource allocation economically attractive.
The resource cost goes up 
when user-space packet processing is introduced or isolating cores, making it
relatively expensive for multiple \vs{} \vm{}s.

\noindent\textbf{Key findings.}
(i) High levels (2x/4x) of virtual network isolation per server can be achieved with an
increase in aggregate throughput (2x) in the shared mode;
(ii) for applications that require low and predictable latency, \vs{}
compartments should use the isolated mode;
(iii) although user-space packet processing using \dpdk{} offers high throughput,
it is expensive (physical \cpu{} and energy costs).

\section{Workload-based Evaluation}
\label{sec:workload}

\begin{figure*}[t!]
	\centering
	\includegraphics[trim=.000cm 0.30cm .35cm 0.25cm,clip=true,height=0.2\textwidth,width=1.00\textwidth]{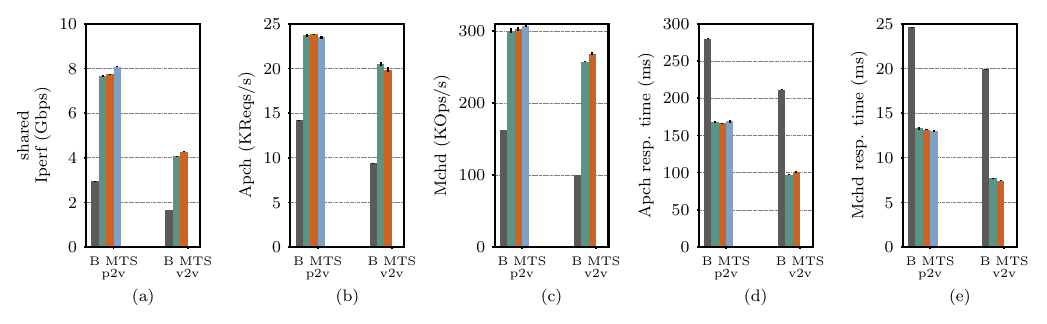}
	\includegraphics[trim=.000cm 0.30cm .35cm 0.25cm,clip=true,height=0.2\textwidth,width=1.00\textwidth]{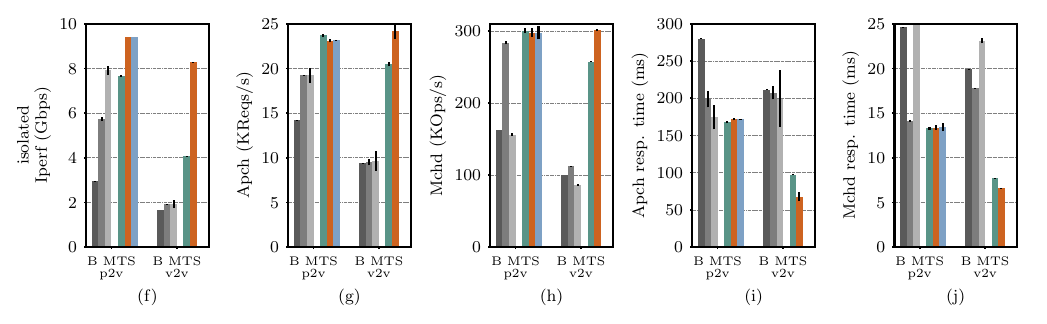}
	\includegraphics[trim=.000cm 0.30cm .35cm 0.25cm,clip=true,height=0.2\textwidth,width=1.00\textwidth]{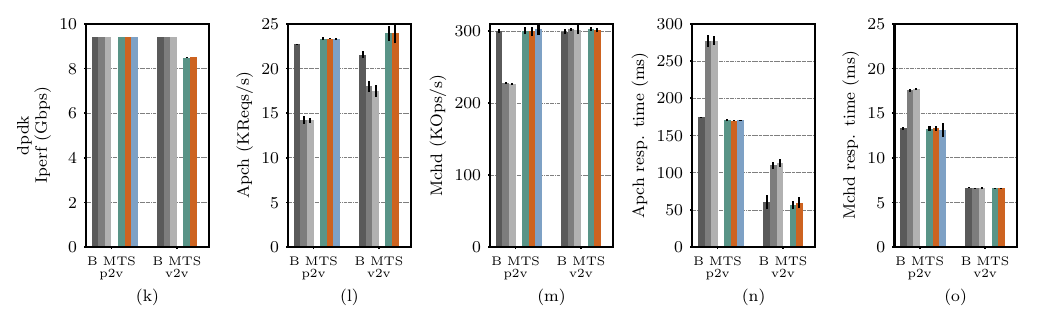}
	\caption{Iperf throughput, \apache{} and \memcached{} throughput and latency
(shown in the columns) comparison of \system.
The rows indicate the resource mode where the bottom row is for security \levelthree{}
in the isolated resource mode combined with the other security levels.
The legend is the same as in Figure~\ref{fig:eval}.}
	\label{fig:workload}
    \vspace{-1.0em}
\end{figure*}

We also conducted experiments with real workloads,
to gain insights on how cloud
applications such as web servers and key-value stores will perform
as tenant applications are the end hosts of the virtual networks.

\noindent\textbf{Methodology. }
For simplicity we focus our workload-based evaluation only on
\tcp{} applications as our previous measurements dealt with \udp{}.
In general, we use a similar methodology to the one described in
Section~\ref{sec:eval}.
For all the \tcp{}-based measurements, we configured the tenant \vm{}s to run the
respective \tcp{} server and from the client (\Lg{})
we benchmark the server to measure the throughput and\slash or response time.
The topologies, resources and setup used to make these measurements are
slightly nuanced which we highlight next.

\noindent\textbf{Traffic scenarios. }
Only the p2v and v2v patterns are evaluated with workloads as we want to
understand the performance of applications hosted in the server.

\noindent\textbf{Resources. }
The ingress and egress ports for all the traffic are on the same physical \nic{}
port unlike in the previous section where the ingress and egress ports were on
separated physical ports of the \nic{}. Hence, each tenant's \vs{} \vm{} was given 1 \vf{} for
In\slash Out and 1 tagged Gw \vf{}. Each tenant \vm{} was given 1 \vf{}.

\noindent\textbf{Setup. }
The applications generating the load are standard \tcp{}, \apache{} and
\memcached{} benchmarking tools respectively Iperf3 v3.0.11~\cite{iperf3}, \apache{}Bench v2.3
(ab)~\cite{ab} and lib\memcached{} v1.0.15 (memslap)~\cite{libmemcached}.
Instead of the Endace card we used a similar Mellanox card at the \Lg{}.

\subsection{Workloads and Results}
\noindent\textbf{Iperf: }
To compare the maximum achievable TCP throughput, we ran Iperf clients for 100
$s$ with a single stream from the \Lg{} to the respective Iperf servers in
the \dut{}'s tenant \vm{}. The aggregate throughput was then reported as the sum of
throughput for each client-server. We collected 5 such measurements for each
experimental configuration and report the mean with 95\% confidence.

\noindent\textbf{Webserver: }
To study workloads from webservers (a very common cloud application),
we consider the open-source \apache{} web
server. Using the \apache{}Bench tool from the \Lg{}, we benchmarked the
respective tenant webservers by requesting a static 11.3 KB web page from four
clients (one for each webserver). Each client made up to 1,000 concurrent connections for 100
$s$ after which we collected the throughput and latency statistics reported by
\apache{}Bench. In the v2v scenario, we used only two client-servers  as one of the
tenant \vm{}s simply forwarded packets using the \dpdk{} $l2fwd$ app.
We collected 5 such repetitions to finally report the average
throughput and latency for each experimental configuration with 95\%
confidence.

\noindent\textbf{Key-value store: }
Key-value stores are also commonly used cloud applications (e.g., with
with webservers). We opted for the open-source \memcached{} key-value store as it
also has an open-source benchmarking tool lib\memcached{}-memslap. We used the
default Set/Get ratio of 90/10 for the measurements. The methodology
and reporting of the measurements are the same as the webserver.

\noindent\textbf{Results. }
The data for the Iperf measurements in the shared mode is shown
Fig.~\ref{fig:workload}(a). The data for the isolated mode is shown in
Fig.~\ref{fig:workload}(f) and Fig.~\ref{fig:workload}(k) depicts the
throughput for \levelthree{}.
As seen in Section~\ref{sec:tput}, here too we observe that \system{} has a higher
throughput (more than 2x in the shared mode) than the \baseline{} except when \dpdk{}
is used in the v2v topology. \system{} saturated the 10G link in the p2v
scenario when isolated and \dpdk{} modes were used.

The data from the throughput measurements for the \apache{} webserver and \memcached{}
key-value store are first reported in the shared mode in
Fig.~\ref{fig:workload}(b) and (c) respectively. For the isolated mode they are shown in
Fig.~\ref{fig:workload}(g) and (h). \levelthree{} throughput is shown in
Fig.~\ref{fig:workload}(l) and (m).
The three main results from the throughput measurements for \apache{} and \memcached{}
are the following. \system{} can offer nearly 2x throughput and 4x isolation
(\leveltwo{}) in the shared mode. \apache{}'s and \memcached{}'s throughput saturated with
\system{}: we expected the throughput to increase as the \vswvm{}s increase when
the compartments have isolated cores, however, we do not observe that. This is
further validated when using \dpdk{}.
\apache{}'s and \memcached{}'s throughput are highly sensitive when the \baseline{}
uses multiple cores in the isolated and \dpdk{} modes which means that using 2 or
more cores requires workload specific tuning to the \host{}: the \dpdk{} parameters,
e.g., drain interval, and the workload \vm{}s, e.g., allocating more cores,
which may not always be necessary with \system{}.

The data from the response time measurements for the \apache{} webserver and \memcached{}
key-value store are first reported in the shared mode in
Fig.~\ref{fig:workload}(d) and (e)~respectively. For the isolated mode they are shown in
Fig.~\ref{fig:workload}(i) and (j). \levelthree{} throughput is shown in
Fig.~\ref{fig:workload}(n) and (o).
Regarding the latency, we again discern that \system{} can offer multiple levels
of isolation and maintain a lower response time (approximately twice as fast) than the \baseline{}.

\noindent\textbf{Key findings. }
Our webserver and key-value store
benchmarks reveal that application throughput and latency of real application
are improved by \system{}. However, for user-space packet processing, the
resource costs go up for a fractional benefit in throughput or latency.  Hence, biting
the bullet for shared resources, offers 4x isolation and approximately 1.5-2x
application performance compared to the \baseline{}.



\section{Discussion}
\label{sec:discussion}

\noindent\textbf{Centralized control, accounting and monitoring. }
\system{} introduces the possibility to realize multi-tenant virtual networks
which can expose tenant/compartment specific interfaces to a logically
centralized control/management plane. This opens up possibilities for full
network virtualization, how to expose the interface, and also how to integrate
\system{} into existing cloud management systems in an easy and usable way.
Furthermore, controllers may need to manage more device, topology and forwarding
information, however, the computations (e.g., routing) should remain the same. From an
accounting and billing perspective, we strongly believe that \system{} is a
new way to bill and monitor virtual networks at granularity more than a simple
flow rule~\cite{gcp-calculator}: \cpu{}, memory and \io{} for virtual networking can be
charged.

\noindent\textbf{\sriov{}: a double-edged sword. }
If an attacker
can compromise \sriov{}, she could violate isolation and in the worst case get
access to the \host{} \os{} via the \pf{} driver. Hence, a rigorous security analysis of
the \sriov{} standard, implementations and \sriov{}-\nic{} drivers can reduce the chance
of a security vulnerability. Compartmentalizing the \pf{} driver is a promising
approach~\cite{boyd2010tolerating}. Furthermore, when a \vswvm{} is shared among
tenants, performance isolation issues could lead to covert
channels~\cite{bates2014detecting} or denial-of-service
attacks~\cite{smolyar2015securing,zhou2017all}. Although not yet
widely supported, \vm{} migration with \sriov{} can be
introduced~\cite{ms-sriov-support}.  \sriov{} \nic{}s have limited \vf{}s
and MAC addresses which could limit the scaling properties of \system{}, e.g.,
when using containers as compartments instead of \vm{}s.

\noindent\textbf{Evaluation limitations.}
The results from our experiments are from a network and application performance
perspective using a 10 Gbps NIC. For a deeper understanding of the performance
improvement we obtained in this paper using SR-IOV, further measurements are
necessary, e.g., using the performance monitoring unit (PMU) to collect a
breakdown of the packet processing latencies. Such an understanding is important
and relevant when dealing with data center applications that require high NIC
bandwidth, e.g., 40/100 Gbps.

As described by 
Neugebauer et al.~\cite{neugebauer2018understanding}, the PCIe bus
can be a bottleneck for special data center applications (e.g., ML
applications): A \emph{typical} x8 PCIe 3.0 NIC (with a maximum payload size of
256 bytes and maximum read request of 4096 bytes) has an effective (usable)
bi-directional bandwidth of approximately 50 Gbps. Hence, the usability of
\system{} with PCIe 3.0 and 8 lanes can indeed be a limitation which we did not
observe in this paper. Nevertheless, increasing the lanes to x16 is one
potential workaround to double the effective bandwidth to around 100 Gbps.
Furthermore, with chip vendors initiating PCIe 4.0 devices~\cite{broadcom-pci4},
the PCIe bus bandwidth will increase to support intense \io{} applications.

\section{Related Work}
\label{sec:relatedwork}

There has been noteworthy research and development on
isolating multi-tenant virtual networks in cloud (datacenter) networks:
tunneling protocols have been standardized~\cite{vxlanrfc,geneverfc},
multi-tenant datacenter architectures have been proposed~\cite{netvirt-mtd}, and real
cloud systems have been built by many companies~\cite{accelnet,andromeda-2018}.
However, most of the previous work still co-locates the \vs{} with the
\host{} as we discussed in Section~\ref{sec:analysis-state-art}. Hence, here
we discuss previous and existing attempts specifically addressing the
security weakness of \vs{}es.

To the best of our knowledge, in 2012 Jin et al.~\cite{181358} (see Research
prototype in Table~\ref{tab:switches}) were the first to point out
the security weakness of co-locating the virtual switch with the hypervisor.
However, the proposed design, while ahead of its time,
(i) lacks a principled approach which this paper
proposes; (ii) has only a single \vs{} \vm{} whereas \system supports multiple
\vs{} compartments making it more robust; (iii) is resource (compute and memory) intensive as the
design used shared memory between the \vs{} \vm{} and all the tenant \vm{}s while
\system uses an inexpensive interrupt-based \sriov{} network card for complete
mediation of tenant-\vs{}-\vm{} and tenant-host networking;~(iv)~requires
considerable effort, expertise and tuning to integrate into
virtualization system whereas \system can easily be scripted into existing cloud
systems.

In 2014 Stecklina~\cite{Stecklina} followed up on this work and proposed sv3, a
user-space switch, which can enable multi-tenant virtual switches (see sv3 in
Table~\ref{tab:switches}). sv3
adopts user-space packet processing and also supports
compartmentalization, i.e., the \host{} can run multiple \vs{}es. However, it
is still co-located with the \host{}, partially adopts the security principles
outlined in this paper, lacks support for \emph{real} cloud
virtual networking, and requires changes to QEMU. Our system on the other hand
moves the \vs{} out of the \host{}, supports production \vs{}es such as \ovs{}
and does not require any changes to QEMU.

Between 2016 and 2017, Panda et al.~\cite{panda2016netbricks}
and Neves et al.~\cite{p4sandbox}
took a language-centric approach to enforce data plane isolation for virtual networks.
However, language-centric approaches require existing \vs{}es to be
reprogrammed\slash annotated which reduces adoption. Hence the solution of using compartments
and \sriov{} in \system allows existing users to easily migrate using their existing
software. Shahbaz et al.~\cite{shahbaz2016pisces} reduced the attack surface of
\ovs by introducing support for the P4 domain specific language which reduces
potentially vulnerable protocol parsing logic.

In 2018, Pettit et al.~\cite{pettit2018bringing} proposed to isolate virtual
switch packet processing using eBPF: which is conceptually isolating potentially
vulnerable parsing code in a kernel-based runtime environment. However, the
design still co-locates the virtual switch and the runtime with the \host{}.

\section{Conclusion}
\label{sec:conclusion}

This paper was motivated by the observation that while
\vs{}es have been designed to enable multi-tenancy, today's \vs{}
designs lack strong isolation between tenant virtual networks. Accordingly,
we presented a novel \vs{} architecture which extends the benefits
of multi-tenancy to the virtual switch, offering improved isolation
and performance, at a modest additional resource cost. When used in
the \emph{shared} mode (only one extra core), with four \vs{} compartments 
the forwarding throughput (in pps) is 1.5-2 times better than the \baseline{}.
The tenant workloads (webserver and key-value stores) we evaluated also receive a
1.5-2 times performance (throughput and response time) improvement with \system{}.

We believe that \system{} is a pragmatic solution that can enhance the security and
performance of virtual networking in the cloud. In particular, \system{}
introduces a way to schedule an entire core for tenant-specific network
virtualization which has three benefits:
(i) application  and packet processing performance is improved;
(ii) this could be integrated into pricing models to appropriately charge customers
on-demand and generate revenue from virtual networking for example;
(iii) virtual network and \host{} isolation is maintained even when the \vs{} is
compromised.

\section{Acknowledgments}
We thank our shepherd Leonid Ryzhyk and our anonymous reviewers for
their valuable feedback and comments. We thank Ben Pfaff, Justin Pettit, Marcel
Winandy, Hagen Woesner, Jean-Pierre Seifert and the Security in
Telecommunications (SecT) team from TU Berlin for valuable discussions at
various stages of this paper.
The first author (K.~T.) acknowledges the
financial support by the Federal Ministry of Education and Research of Germany
in the framework of the Software Campus 2.0 project nos.~01IS17052 and 01IS1705,
and the ``API Assistant'' activity of EIT Digital.
The third author (G.~R.) is with the 
MTA-BME
Information Systems Research Group.

{
\balance
{\bibliographystyle{acm}
\bibliography{master}}
\balance
}





\end{document}

%% file: tables/switches.tex
\begin{table}[t]
\normalsize
\caption{Design characteristics of virtual switches.}
\label{tab:switches}
\scriptsize
\centering
\begin{tabular}{@{}p{2.0cm}p{0.3cm}p{0.4cm}p{0.05cm}p{1.5cm}p{0.1cm}p{0.1cm}p{0.1cm}p{0.1cm}@{}}
\textbf{Name} & \textbf{Ref.} & \textbf{Year} &  & \textbf{Emphasis} & \rotatebox{45}{\textbf{Monolithic}} &\rotatebox{45}{\textbf{Co-Location}} & \rotatebox{45}{\textbf{Kernel}} & \rotatebox{45}{\textbf{User}} \\ \bottomrule
& & & & & & & & \\
\ovs & \cite{ovsStart} & 2009 &  &  Flexibility & \checkmark & \checkmark & \checkmark & \checkmark \\
\rowcolor{Gray}
\cellcolor{white}
Cisco NexusV & \cite{nexusVStart} & 2009  &  & Flexibility & \checkmark & \checkmark & \checkmark & \xmark \\
VMware vSwitch & \cite{vmwareStart} & 2009  &  & Centralized control & \checkmark & \checkmark & \checkmark & \xmark \\
\rowcolor{Gray}
\cellcolor{white}
Vale & \cite{Rizzo:2012:VSE:2413176.2413185} & 2012 &  & Performance & \checkmark & \checkmark & \checkmark & \xmark \\
Research prototype & \cite{181358} & 2012  &  & Isolation & \checkmark & \xmark & \checkmark & \checkmark \\
\rowcolor{Gray}
\cellcolor{white}
Hyper-Switch & \cite{ram2013hyper} & 2013 &  & Performance & \checkmark & \checkmark & \checkmark & \checkmark \\
MS HyperV-Switch & \cite{msvSwitch} & 2013  &  & Centralized control & \checkmark & \checkmark & \checkmark & \xmark \\
\rowcolor{Gray}
\cellcolor{white}
NetVM & \cite{hwang2014netvm} & 2014 &  & Performance, NFV & \checkmark & \checkmark & \xmark & \checkmark \\
sv3 & \cite{Stecklina} & 2014 &  & Security & \xmark & \checkmark & \xmark & \checkmark \\
\rowcolor{Gray}
\cellcolor{white}
fd.io & \cite{fdio} & 2015 & & Performance & \checkmark & \checkmark& \xmark & \checkmark \\
mSwitch & \cite{honda2015mswitch} & 2015 & & Performance & \checkmark & \checkmark & \checkmark & \xmark \\ 
\rowcolor{Gray}
\cellcolor{white}
BESS & \cite{bess} & 2015  &  & Programmability, NFV & \checkmark & \checkmark & \xmark & \checkmark \\
PISCES & \cite{shahbaz2016pisces} & 2016 & & Programmability & \checkmark & \checkmark & \checkmark & \checkmark \\
\rowcolor{Gray}
\cellcolor{white}
\ovs with DPDK & \cite{dpdkOvs} & 2016 &  &  Performance & \checkmark & \checkmark & \xmark & \checkmark \\
ESwitch & \cite{molnar2016dataplane} & 2016 & & Performance& \checkmark & \checkmark & \xmark & \checkmark \\
\rowcolor{Gray}
\cellcolor{white}
MS VFP & \cite{firestone2017vfp} & 2017 & & Performance, flexibility & \checkmark & \checkmark & \checkmark & \xmark \\

Mellanox BlueField & \cite{mellanox-bluefield} & 2017 & & CPU offload & \checkmark & \xmark & \checkmark & \checkmark \\
\rowcolor{Gray}
\cellcolor{white} 
Liquid IO & \cite{smartnic-cavium} & 2017 &  &  CPU offload & \checkmark & \xmark & \checkmark & \checkmark \\
Stingray & \cite{smartnic-broadcom} & 2017 &  &  CPU offload & \checkmark & \xmark & \checkmark & \checkmark \\
\rowcolor{Gray}
\cellcolor{white} 
GPU-based \ovs{} & \cite{ovs-gpu} & 2017 &  &  Acceleration & \checkmark & \checkmark & \checkmark & \checkmark \\
MS AccelNet & \cite{accelnet} & 2018 &  &  Performance, flexibility& \checkmark & \checkmark & \checkmark & \xmark \\
\rowcolor{Gray}
\cellcolor{white} 
Google Andromeda & \cite{andromeda-2018} & 2018 &  & Flexibility and performance& \checkmark & \checkmark & \xmark & \checkmark \\\midrule
\end{tabular}
\vspace{-1.0em}
\end{table}
